\documentclass[journal,lettersize,onecolumn]{IEEEtran}
\IEEEoverridecommandlockouts
\usepackage{cite}
\usepackage{amsmath,amsthm,amssymb,amsfonts}
\usepackage{algorithmic}
\usepackage{graphicx}
\usepackage{epstopdf}
\usepackage{textcomp}
\usepackage{xcolor}
\usepackage{balance}
\usepackage{tikz}
\usepackage{matlab-prettifier}
\usepackage{multicol}
\usepackage{hyperref}
\usepackage{blindtext}
\usepackage{float}
\hypersetup{
	colorlinks=true,
	linkcolor=blue,
	anchorcolor=red,
	filecolor=blue,      
	urlcolor=blue,
	pdftitle={Mediumband By D A Basnayaka},
	pdfpagemode=FullScreen,
	frenchlinks=false,
	citecolor=blue,
	citebordercolor=blue,
	pdfauthor=D A Basnayaka,
}

\newcommand{\BE}{\begin{equation}}
	\newcommand{\EE}{\end{equation}}
\newcommand{\bm}[1]{\mbox{\boldmath $#1$}}

\newcommand{\var}{{\rm var}}

\def\BibTeX{{\rm B\kern-.05em{\sc i\kern-.025em b}\kern-.08em
		T\kern-.1667em\lower.7ex\hbox{E}\kern-.125emX}}
\begin{document}
	\title{Simulating Mediumband Wireless Communication Systems: A Concise Description
	\thanks{D. A. Basnayaka is a Visiting Professor at School of Engineering,
		ESIGELEC, Rouen, France. This paper includes comprehensively-commented MATLAB codes for the sake of verifiability and reproducibility, and the readers may contact the corresponding author for more information, or comments, or corrections. Corresponding Author: d.basnayaka@ieee.org. 
	}} 
	\author{\centering{Dushyantha A Basnayaka, \textit{Senior Member IEEE}} \\
	\vspace{5mm}
	\textit{[Tutorial Paper with MATLAB Codes]}
	\vspace{-4mm}}
	\maketitle
	\begin{abstract}
		In this paper, we describe the necessary procedures for accurately simulating digital wireless communication systems operating in the mediumband, aimed at both beginners and experts. In the research literature, digital wireless communication systems are typically simulated in the discrete-time complex baseband domain, where pulse shaping, upconversion, mixing, carrier synchronization, and symbol timing synchronization are often ignored. These assumptions are indeed sufficient in most cases, but to capture the essence of communication in the mediumband, certain physical layer (PHY) operations should be simulated in detail. In this paper, we concisely describe how to simulate a mediumband wireless communication scenario—from a single transmitter (TX) to a single receiver (RX)—in MATLAB, elaborating the operation of key PHY subsystems. The approach described here ensures that the simulated system captures the delicate dynamics of mediumband wireless communication, including the effect of deep fading avoidance.  
	\end{abstract}
	\begin{IEEEkeywords}
		Mediumband, multipath, delay spread, symbol period, pulse shaping, AWGN \end{IEEEkeywords}
	\vspace{0mm}
	\tableofcontents
	%
%	\vspace{0mm}
	%
	%
\section{Introduction}
Mediumband wireless communication refers to digital wireless communication that satisfies a set of constraints defined in terms of delay spread and symbol period \cite{Bas2023}. The term “mediumband” does NOT refer to any particular band of frequencies in the electromagnetic (EM) spectrum like microwave band, mmwave band or terahertz band. Furthermore, mediumband conceptually has nothing to do with another closely sounding term: mid-band, which just refers to a band of frequencies in the EM spectrum roughly from 3.5GHz to 25GHz. It has been shown that wireless communication systems that satisfy those mediumband constraints possess many unique properties that could be instrumental in the design of high-rate and highly reliable wireless communication systems\cite{BasMag2024}.\\
\indent For future research, accurately simulating mediumband wireless communication systems is of paramount importance. Typically the de-facto standard for wireless communication system simulation is the simulation technique that focuses on discrete-time baseband equivalent signals assuming perfect pulse shaping, upconversion, RF mixing, filtering and carrier synchronization, and symbol timing synchronization. However, to observe the true impact of operating in the mediumband, some of the operations listed above should be considered in the simulations.\\  
\indent In this concise note, elaborating key subsystems in a typical digital wireless communication system, we describe how mediumband wireless communication systems should be properly simulated to capture the key dynamics of the fading, input and noise processes. In particular, we highlight the need of considering baseband equivalent signals in analog form, and the importance of accurate symbol timing synchronization and optimal sampling at the receiver. Without accurate symbol timing synchronization and optimal sampling, some of the key effects that are unique to mediumband wireless communication like the effect of deep fading avoidance would not be prominent and could not be studied.\\
\indent For brevity, we omit detailed description on mediumband wireless communication and its potential impact for wider wireless communication in this paper, but interested readers are referred to \cite{Bas2023,BasMag2024,BasFirag25} for more details.  
\begin{figure*}[t]
	\centerline{\includegraphics[scale=0.82]{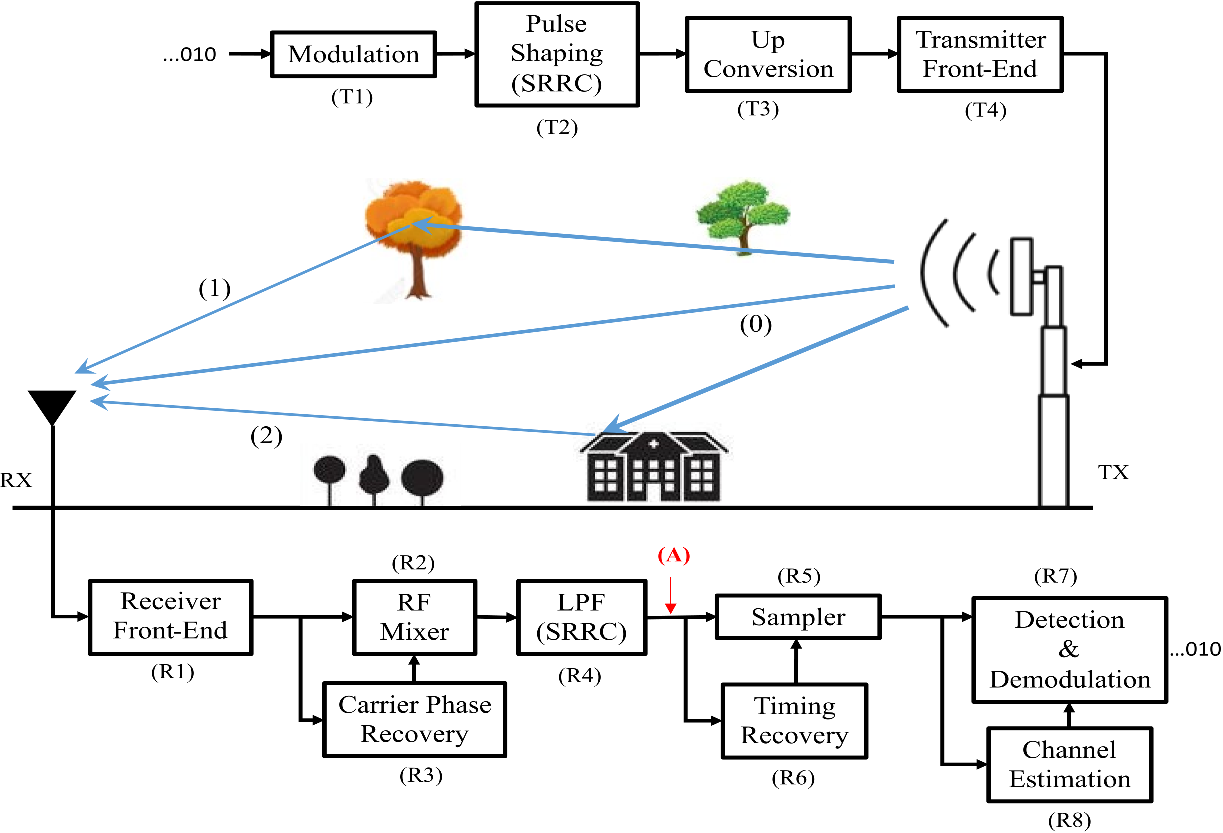}}
	\caption{Typical sub-systems of a modern digital wireless communication scenario between a single transmitter and a single receiver.}\label{fig:fig1}
\end{figure*}  
%
%\vspace{-5mm}
%
\section{System Model}
%
%
%\vspace{-7mm}
%
We consider an uncoded wireless communication scenario in which a single transmitter (TX) and a single receiver (RX), as shown in Fig. \ref{fig:fig1}, operate in a rich scattering environment. At the physical layer (PHY), a typical TX and RX consist of many key sub-systems.\\
\indent The TX includes sub-systems for modulation, pulse shaping, and upconversion. The modulation unit, in conjunction with a signal constellation known to both TX and RX, modulates the digital data corresponding to a single PHY frame into a complex vector. Typically, prior to modulation, coding may be applied, but it is omitted here as it falls outside the scope of this concise discussion \cite{Ungerboeck82}. Pulse shaping is subsequently applied to the modulated complex data separately on the I and Q branches using square-root-raised-cosine (SRRC) filters. The pulse-shaped signal is then upconverted using a suitable carrier wave before being transmitted from the transmitting antenna.\\
\indent The transmit signal, after propagating through the propagation environment, reaches the RX. As a result of scattering, the received signal is typically a mixture of delayed and attenuated versions of the same transmit signal. After passing through a low-noise amplifier (LNA) in (R1), the signal mixture is mixed with a local oscillator (LO) and then low-pass filtered using a matching SRRC filter as depicted by (R2) and (R4) respectively as shown in Fig. \ref{fig:fig1}.\\
\indent Our detailed discussion starts considering the analog complex baseband equivalent receive signal corresponding to a single PHY frame at (A) in Fig. \ref{fig:fig1}, which is:
\begin{align} \label{eq00}
	r(t) &= \sqrt{E_s}\sum_{n=0}^{N-1} \alpha_n e^{-\phi_n} s(t-\tau_n) + w(t),
\end{align}
where $\alpha_n$, $\phi_n$, and $\tau_n$ are respectively known as the amplitude, phase and the propagation delay of the $n$th multipath component (MPC). For notational simplicity, let $\gamma_n=\alpha_n e^{-\phi_n}$, and $y(t)$ be the receive signal in the absence of noise, so: 
\begin{align} \label{eq0}
	r(t) &= \overbrace{\sqrt{E_s}\sum_{n=0}^{N-1} \gamma_n s(t-\tau_n)}^{y(t)} + w(t).
\end{align}
The propagation parameters, $\tau_n$ and $\gamma_n$ are assumed to be fixed at least within the time duration of a single frame. The signal, $s(t)$ is the normalized data signal corresponding to a single PHY frame (see Fig. \ref{fig:fig11}-(a))):
\begin{figure*}[t]
	\centerline{\includegraphics[scale=0.8]{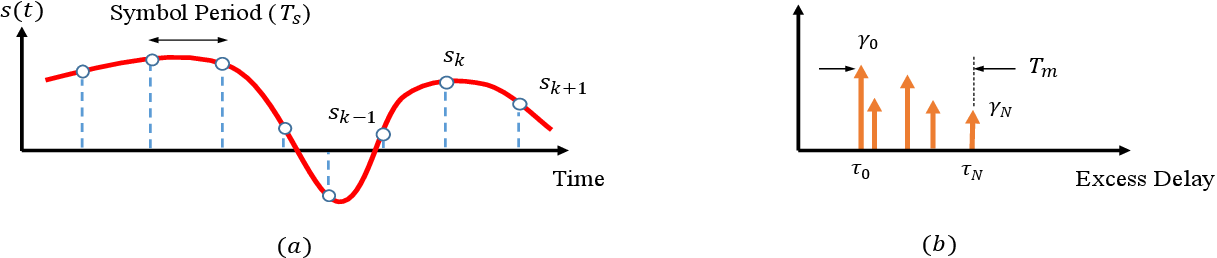}}
	\caption{a) Typical data signal $s(t)$, where only the data points labelled in $s_k$ should be estimated at the RX, and b) Typical channel impulse response (CIR) caused by multipath propagation, where the delay spread $T_m$ is the time difference between the earliest and the latest major MPCs.}\label{fig:fig11}
\end{figure*} 
\begin{align}\label{eq1}
	s(t) &= \sum_k s_k p(t-kT_s),
\end{align}
where $T_s$ is the symbol period, and $\{s_0, s_1\dots, s_k, \dots, s_{L-1}\}$ is the sequence of symbols drawn from a constellation, $\mathcal{Q}$ of size $M$ (e.g. BPSK, 4QAM) meaning $s_k \in \mathcal{Q}$. The $L$ is the length of the PHY frame in terms of symbol periods. For BPSK, $M=2$ and for 4QAM $M=4$, and the symbols are normalized such that $\mathcal{E}\left\{|s_k|^2\right\}=1$. The constant factor, $E_s$ captures the effect of the transmit power on the receive signal $r(t)$, and the $w(t)$ is the filtered complex noise process.\\
\indent In $p(t)$ in \eqref{eq1}, we incorporate the effect of matching SRRC filtering at the TX and RX, and assume the effect of upconversion, TX and RX front-ends, carrier phase recovery, and RF mixing are perfect. The combine effect of TX and RX pulse shaping yields the raised-cosine (RC) filter for $p(t)$:
\begin{align}
	p(t) &= p_T(t) \circledast p_R(t),
\end{align}
where $p_T(t)$ and $p_R(t)$ are the SRRC pulse shaping filters applied at (T2) and (R4) respectively. Here $\circledast$ is the convolution operator, and $p(t)$ is:
\begin{align}\label{eq2}
	p(t) &= \begin{cases}
		\frac{\pi}{4}\textup{sinc}\left(\frac{1}{2\beta}\right), & t = \pm \frac{T_s}{2\beta} \\
		\textup{sinc}\left(\frac{t}{T_s}\right)\frac{\cos\left(\frac{\pi \beta t}{T_s}\right)}{1-\left(\frac{2\beta t}{T_s}\right)^2}, & \left|t\right| \leq KT_s, \\
		0, & \text{otherwise.}
	\end{cases}
\end{align}
where $\beta$ is the roll-off factor. The integer, $K$ is such that $K \in \left[2,6\right]$, which means the span of the pulse $p(t)$ is limited to $2K$ symbol periods. In modern wireless communication, $p_T(t)$ and $p_R(t)$ are such that their Fourier transforms are: \cite{Robert19}:
\begin{subequations} \label{eq201}
	\begin{align}
		P_T(f) &= \sqrt{T_s P(f)}, \\
		P_R(f) &= \sqrt{\frac{P(f)}{T_s}},
	\end{align}
\end{subequations}
%t
where $P(f)=\mathcal{F}\left\{p(t)\right\}$ is the Fourier transform of the RC pulse $p(t)$, and $P(f)$ is in \eqref{eq2n}.
\begin{figure*}[t]
\begin{align}\label{eq2n}
	P(f) &= \begin{cases}
		T_s, & 0 \leq |f| \leq \frac{1-\beta}{2T_s} \\
		\frac{T_s}{2}\left(1+\cos\frac{\pi T_s}{\beta}\left(f-\frac{1-\beta}{2T_s}\right)\right), & \frac{1-\beta}{2T_s} \leq |f| \leq \frac{1+\beta}{2T_s}, \\
		0 & \text{otherwise.}
	\end{cases}
\end{align}
\hrule
\end{figure*}
Note that the $P(f)$ in \eqref{eq2n} is such that
\begin{align}\label{eq2n1}
\int_{-\infty}^{\infty} P(f) df = 1.
\end{align} 
\begin{figure*}[t]
	\begin{align}\label{eq:srrc}
		\Phi(t) &= \begin{cases}
			1+\beta\left(\frac{4}{\pi}-1\right), & t = 0 \\
			\frac{\beta}{\sqrt{2}}\left[\left(1+\frac{2}{\pi}\right)\textup{sin}\left(\frac{\pi}{4\beta}\right)+\left(1-\frac{2}{\pi}\right)\textup{cos}\left(\frac{\pi}{4\beta}\right)\right], & t =\pm \frac{T_s}{4\beta} \\
			\frac{\sqrt{T_s}}{\pi t}\left[\frac{\sin\left(\frac{\pi t(1-\beta)}{T_s}\right)+\frac{4\beta t}{T_s}\cos\left(\frac{\pi t(1+\beta)}{T_s}\right)}{1-\left(\frac{4\beta t}{T_s}\right)^2} \right], & \text{otherwise.}
		\end{cases}
	\end{align}
	\hrule
\end{figure*}    
Furthermore, if needed, one can find the time domain representation of $p_T(t)$ and $p_R(t)$ as $p_T(t)=\sqrt{T_s}\Phi(t)$ and $p_R(t)=\frac{\Phi(t)}{\sqrt{T_s}}$, where $\Phi(t)$ is in \eqref{eq:srrc}, which in fact is:
\begin{align}
	\setcounter{equation}{9}
	\Phi(t)&=\mathcal{F}^{-1}\left\{\sqrt{P(f)}\right\}.
\end{align}
\indent Furthermore, as shown in Fig. \ref{fig:fig11}-(b), let $T_m$ be the delay spread defined as the time difference between the earliest and the latest major MPCs, which mathematically is:
\begin{align}
	T_m &= \max_{n} \left| \tau_n - \tau_0\right|.
\end{align}
\indent The RX subsequently samples the complex signal, $r(t)$ for I and Q branches separately at regular intervals to obtain the complex vector, $\bm{r}=\left\{r(0), r(1), \dots, r(k), \dots, r(L-1)\right\}^T$. Let $t=\hat{\tau}+kT_s$ be the time instances when the I and Q branches are sampled. So:
\begin{align}\label{eq3}
	r(k) &= \left.r(t)\right|_{t=\hat{\tau}+kT_s},
\end{align}
for $k=0,1 \dots, L-1$, where $\hat{\tau}$ is the symbol timing offset, which should be estimated prior to sampling. In mediumband wireless communication system simulations, the symbol timing synchronization in (R6), otherwise finding this optimum timing offset, is very important. Otherwise, the potential gain achievable by communicating in the mediumband is limited. The Sec. \ref{subsec:symbol_timing} elaborates the symbol timing synchronization.\\
\indent Exploiting the receive sample vector $\bm{r}$ and the estimated channel state information (CSI), the detection and demodulation unit, which is (R7), estimates the data at the RX.
\subsection{Mediumband Constraint}\label{sec:mediumband}
\indent When a given digital wireless communication system satisfies the constraint \cite{Bas2023}:
\begin{align}
	\text{\textbf{Mediumband}:} \qquad \quad \quad  T_m < T_s < 10T_m,
\end{align}
the corresponding wireless system is said to be operating in the mediumband or known as a mediumband wireless communication system, where, as shown in Fig. \ref{fig:fig11}-(b), $T_m$ is the delay spread.\footnote{Similarly, when $T_m \leq 0.1T_s$, or colloquially $T_m \ll T_s$, the wireless system is said to be operating in the narrowband, and when $T_m > T_s$, the wireless system is said to be operating in the broadband. See Table 1.}. As we call it, the degree of mediumbandness of mediumband wireless communication is measured in terms of percentage delay spread (PDS), which is defined as:
\begin{align}\label{eq:pds1}
	\text{PDS} &= \left(\frac{T_m}{T_s}\right) \text{ x } 100 \%.
\end{align}
Why is this particular mediumband constraint so special, and why does it matter for future wireless communication? These are important questions, but are outside the scope of this paper. However, in summary, when systems operate in the mediumband, they can intelligently weight the multipath components in such a way that the deep fading in the desired fading factor, which is $g^{\bullet}$ in \eqref{channel:DT:eq1}, is significantly reduced even in highly non-line-of-sight (NLoS) propagation\footnote{The probability density function (PDF) of the desired fading factor in NLoS propagation has a peak at zero, but when the systems operate in the mediumband, the PDF of the desired fading factor has a trench instead at zero meaning the probability of the desired fading factor being small is low.}.  This effect in the PDF is important because, in wireless communication, it is the statistics of the fading factors, not the specific knowledge of them, that matter more. The readers are referred to the key references for more information \cite{Bas2023,BasMag2024,BasFirag25}. 
\begin{figure*}
	\setcounter{equation}{15}
	\begin{align} \label{auto-corr:eq1}
		R_{ss}\left(\tau\right) = \left.\operatorname{sinc}\left( \frac{\tau}{T_s} \right) \frac{\cos\left( \beta \frac{\pi \tau}{T_s} \right)}{1 - \left( \frac{2 \beta \tau}{T_s} \right)^2} - \frac{\beta}{4} \operatorname{sinc}\left(\beta \frac{\tau}{T_s} \right) \frac{\cos\left( \frac{\pi \tau}{T_s} \right)}{1 - \left( \frac{\beta \tau}{T_s} \right)^2} \right.
	\end{align}
	\hrule
\end{figure*} 
\begin{table}[t]
	\begin{center}
		\caption{Mediumband Constraint As Defined in \cite{Bas2023}}
		\label{table:1}
		\begin{tabular}{|c|c|} 
			\hline
			Class & Relationship \\
			\hline
			Narrowband & $T_m \ll T_s $ \\
			\hline 
			Mediumband & $T_m < T_s < 10T_m$ \\
			\hline 
			Broadband & $T_s \ll T_m$ \\ 
			\hline
		\end{tabular}
	\end{center}
	\vspace{-3mm}
\end{table} 
\subsection{Symbol Timing Synchronization}\label{subsec:symbol_timing}
To perform the sampling in \eqref{eq3} optimally, the optimum timing offset, $\hat{\tau}$, is required. It can be determined via exhaustive search, which basically solves the optimization \cite{Firag24}:
\begin{align}\label{eq4}
	\setcounter{equation}{14}
	\hat{\tau} &= \max_t \left\{\left|\sum_{n=0}^{N-1} \gamma_n R_{ss}(\tau_n-t)\right|^2 \right\},
\end{align}
where $R_{ss}(\tau)=\mathcal{E}\left\{s(t)\overline{s(t+\tau)}\right\}$ is the autocorrelation function of $s(t)$, where the `overline' denotes the complex conjugation, and $\gamma_n$ is the complex channel gain of the $n$th MPC. We can show that $R_{ss}(\tau)$ is equal to \eqref{auto-corr:eq1} \cite{Bas2023}. The optimization in \eqref{eq4} effectively searches the excess delay axis to find the time instance that gives rise to the largest objective function. The objective function in \eqref{eq4} is in fact functionally related to the desired fading factor in mediumband channels as originally shown in \cite[Theorem 1]{Bas2023}. Also see equation \eqref{eq:H0}.\\
\indent Furthermore, if $s(t)$ is a real signal, applying symbol timing synchronization and sampling to I and Q branches separately would be optimal. If that is the case, the following optimizations:
\setcounter{equation}{16}  
\begin{align}
	\hat{\tau}_i &= \max_t \left\{\left|\sum_{n=0}^{N-1} \text{Re}\left\{\gamma_n\right\} R_{ss}(\tau_n-t)\right|^2\right\}, \label{eq:toi} \\
	\hat{\tau}_q &= \max_t \left\{\left|\sum_{n=0}^{N-1} \text{Im}\left\{\gamma_n\right\} R_{ss}(\tau_n-t)\right|^2 \right\},
	\label{eq:toq}
\end{align}
%\end{subequations}
%
should be solved, where $t=\hat{\tau}_i+kT_s$ and $t=\hat{\tau}_q+kT_s$ are the time instances where I and Q branches are sampled respectively. Here $\text{Re}\left\{\gamma_n\right\}$ and $\text{Im}\left\{\gamma_n\right\}$ are respectively the real and imaginary parts of $\gamma_n$. Consequently $k$th sample of $r(t)$ would be:
\begin{subequations}\label{eq:samplingIQ}
\begin{align}
	\text{Re}\left\{r(k)\right\} &=\left.r_i(t)\right|_{t=\hat{\tau}_i+kT_s}, \\
	\text{Im}\left\{r(k)\right\} &=\left.r_q(t)\right|_{t=\hat{\tau}_q+kT_s},
\end{align}
\end{subequations}
for $k=0,1 \dots, L-1$, where $r_i(t)$ and $r_q(t)$ are the baseband equivalent received signals by I and Q branches respectively at the RX, and $r(t)=r_i(t)+jr_q(t)$.
\section{Discrete-Time Mediumband Channel Models}\label{sec:models}
\subsection{Matrix Model}\label{sec:matrix_models}
Assuming the span of the pulse shaping filter is limited to $2K$ symbol periods, the sampled version of $r(t)$, that is $r(0), r(1), \dots, r(L-1)$ corresponding to a single frame can be given in matrix form. Stacking received samples vertically, we can formulate a matrix equation \cite{Firag24}:
\begin{align}
	\bm{r} &=\sqrt{E_s}\bm{H}\bm{s}+\bm{w},
	\label{eq:bigrr}
\end{align}
where $\bm{r}=\{r(0),r(1), ... ,r(L-1)\}^T$, $\bm{s}$ is the ${L}\times1$ transmitted symbols, $\{s_0, s_1, ... ,s_{L-1}\}^T$, and $\bm{w}=\{w(0),w(1), ... ,w(L-1)\}^T$ is the ${L}\times1$ sampled noise vector. We assumed ${L} > {K}$ here, where the  pulse shaping filter has duration of ${2KT_s}$ seconds. The ${L}\times {L}$ channel gain matrix has Toeplitz form, and is:
\begin{equation} \!\!\! \!\!\!\label{eq:HMIMO} \bm{H}  \! \! = \! \!  \left[ \begin{array}{cccccc}
		\! \! \! h_{0}^{\bullet}&h_{1}& h_{2}&\ldots&\ldots&h_{L-1}\! \! \!\\
		\! \! \! h_{-1}&h_{0}^{\bullet}& h_{1}&\ddots&\ldots&\vdots \! \! \! \\
		\! \! \! h_{-2}&h_{-1}& \ddots&\ddots&\ddots&\vdots \! \! \! \\
		\! \! \! \vdots&\ddots& \ddots&\ddots&h_{1}&h_{2}\! \! \! \\
		\! \! \! \vdots&\vdots& \ddots&h_{-1}&h_{0}^{\bullet}&h_{1} \! \! \! \\
		\! \! \! h_{-L+1}&\ldots& \ldots&h_{-2}&h_{-1}&h_{0}^{\bullet} \! \! \! \\
	\end{array} \right]\! \! \!, \! \! \!
\end{equation}
where, as shown briefly in Appendix \ref{App:A}, $h_{\nu}$ is given by:
\begin{align}\label{eq:pdf0}
	h_{\nu} &= \begin{cases}
		\sum_{n=0}^{N-1} \gamma_n p(\hat{\tau}-\tau_n-\nu T_s) &  \nu =-K,...,K,\\
		0 & \text{Otherwise}.
	\end{cases}
\end{align}
Here\footnote{Note that it is only the fading coefficients marked with a ``\textit{filled-black-circle}'' that exhibit the effect of deep fading avoidance. This effect of deep fading avoidance is the effect that gives rise to a trench in the PDF of the corresponding fading coefficient of mediumband wireless communication systems even in NLoS propagation \cite[Fig. 9]{BasMag2024}.}  $h_{0}^{\bullet}=h_{0}$. Note that $\bm{H}$ has only $2K+1$ non-zero elements. For instance, for $K=1$, 
\begin{equation} \!\!\! \!\!\!\label{eq:HMIMOK1} \bm{H}  \! \! = \! \!  \left[ \begin{array}{cccccc}
		\! \! \! h_{0}^{\bullet}&h_{1}& 0 &\ldots&\ldots& 0\! \! \!\\
		\! \! \! h_{-1}&h_{0}^{\bullet}& h_{1}&\ddots&\ldots&\vdots \! \! \! \\
		\! \! \! 0 &h_{-1}& \ddots&\ddots&\ddots&\vdots \! \! \! \\
		\! \! \! \vdots&\ddots& \ddots&\ddots&h_{1}&0\! \! \! \\
		\! \! \! \vdots&\vdots& \ddots&h_{-1}&h_{0}^{\bullet}&h_{1} \! \! \! \\
		\! \! \! 0 &\ldots& \ldots& 0 &h_{-1}&h_{0}^{\bullet} \! \! \! \\
	\end{array} \right]\! \! \!, \! \! \!
\end{equation}
where, one may notice that there are only three unique non-zero elements in \eqref{eq:HMIMOK1}. Similarly, for $K=2$, there would be only 5 non-zero elements in \eqref{eq:HMIMO}. So in general, \eqref{eq:HMIMO} has only $2K+1$ non-zero elements, where $2K$ is the duration of the pulse shaping filters in terms of symbols periods.\\
\indent The independent, identically distributed (iid) complex noise samples are: $w(k) \sim \mathcal{CN}\left(0,\sigma^2\right)$, which effectively means $w_i(k)\sim \mathcal{N}\left(0,\frac{\sigma^2}{2}\right)$ and $w_q(k) \sim \mathcal{N}\left(0,\frac{\sigma^2}{2}\right)$, where $w(k)=w_i(k)+jw_q(k)$. As shown in Appendix \ref{App:B}, $\sigma^2=\frac{N_0}{T_s}$, where $N_0=-174$dBm/Hz at room temperature. In simulations, the $\sigma^2$ is set to satisfy a certain received signal-to-noise ratio $(\text{SNR})$ target. More details on that can be found in Sec. \ref{sec:SNR}.
\subsection{Linear Model}\label{sec:linear_models}
If the symbol timing recovery at the RX is assumed to have been done as described in Sec. \ref{subsec:symbol_timing} and is perfect, a discrete-time model for \eqref{eq00} can be obtained as \cite[eq. 13]{Bas2023}:
\begin{align}\label{channel:DT:eq1}
	r(k) &= \sqrt{E_s}g^{\bullet}s_k + \mathcal{I}(k) + w(k),
\end{align}
for $k=0,1, \dots, L-1$, where $\sqrt{E_s}g^{\bullet}s_k$, $\mathcal{I}(k)$, $w(k)$ are respectively the desired, residual interference, and the noise samples. The $g^{\bullet}$ is the desired fading factor, which is \cite[Theorem 1]{Bas2023}:
\begin{align}\label{eq:H0}
	g^{\bullet} &= \frac{\sum_{n=0}^{N-1} \gamma_n R_{ss}(\tau_n-\hat{\tau})}{1-\frac{\beta}{4}}.
\end{align}
If I and Q branches are synchronized  and sampled separately as described in \eqref{eq:toi}, \eqref{eq:toq} and \eqref{eq:samplingIQ}, the desired fading factor would be:
\begin{align}\label{eq:H}
	\text{Re}\left\{g^{\bullet}\right\} &= \frac{\sum_{n=0}^{N-1} \text{Re}\left\{\gamma_n\right\} R_{ss}(\tau_n-\hat{\tau}_i)}{1-\frac{\beta}{4}},\\
	\text{Im}\left\{g^{\bullet}\right\} &= \frac{\sum_{n=0}^{N-1} \text{Im}\left\{\gamma_n\right\} R_{ss}(\tau_n-\hat{\tau}_q)}{1-\frac{\beta}{4}},
\end{align}
where $R_{ss}(\tau)$ is in \eqref{auto-corr:eq1}.
\subsection{Signal-To-Noise Ratio (SNR)} \label{sec:SNR}
The average power of the information signal to the noise signal in \eqref{eq0} can be shown to be equal to:
\begin{align}
	 \frac{\mathcal{E}\left\{\left|y(k)\right|^2\right\}}{\mathcal{E}\left\{\left|w(k)\right|^2\right\}}&=\frac{E_s}{\sigma^2}\left(\sum_{n=0}^{N-1}\mathcal{E}\left\{|\gamma_n|^2\right\}\right) = \frac{E_sT_s\Gamma}{N_0},\label{eq:snr0}
\end{align}
where $y(k)$s and $w(k)$s are the symbol-spaced samples of $y(t)$ and $w(t)$ respectively in \eqref{eq0} taken at $t=\hat{\tau}+kT_s$, and $\Gamma=\left.\sum_{n=0}^{N-1}\mathcal{E}\left\{|\gamma_n|^2\right\}\right.$. In wireless communication, SNR is defined as the signal energy per bit to $N_0$ \cite[Eq. 6.1]{Gold05}. By noting the fact that the numerator in \eqref{eq:snr0} is the signal energy per symbol period, and each symbol represents $\log_2M$ bits of information, the receive SNR would be:      
\begin{align}
\text{SNR} &\stackrel{\Delta}{=} \frac{E_sT_s\Gamma}{N_0 \log_2M} = \frac{E_s\Gamma}{\sigma^2 \log_2M}.
\end{align}
\indent In order to set the simulation to a certain receive SNR, with the knowledge of $E_s$ and $\Gamma$, the $\sigma^2$ should be set to:
\begin{align}
	\sigma^2 &= \frac{E_s\Gamma}{\text{SNR}\log_2M}.
\end{align}
To simplify the simulation, setting $\Gamma$ and $E_s$ to unity is typical. As a result, for BPSK, $\sigma^2=1/\text{SNR}$. Alternatively one can choose $\sigma^2=1$, and $E_s$ for $y(t)$ in \eqref{eq0} can be set to satisfy a certain $\text{SNR}$ requirement as well as $E_s=\sigma^2\text{SNR}$.        
\section{Performance Analysis}\label{sec:ber}
The common practice for evaluating the performance of wireless communication systems is to consider the average uncoded bit-error-rate (BER) over a range of SNR values, typically from -5dB to 30dB. We evaluate that of mediumband communication systems using the receive sample vector, $\bm{r}$ and models in Sec. \ref{sec:models}. For instance, the matrix model in \eqref{eq:bigrr} can be used to find the MMSE estimate of $\bm{s}$: $\hat{\bm{s}}$ as:
\begin{align}\label{nb:statistics:eq1}
\bm{\hat{s}}=\bm{W}\bm{r},
\end{align} 
where the ${L}\times {L}$ MMSE processing matrix: $\bm{W}$ can be found straightforwardly as: $\bm{W}=\bm{H}^\dagger\left(\bm{H}\bm{H}^\dagger+\frac{\sigma^2}{E_s}\bm{I}\right)^{-1}$ \cite[Eq. 5.28]{Ezio07}. Here $(.)^\dagger$ is the conjugate transpose operation. In this particular case, we assume RX has the complete knowledge of $\bm{H}$. In the absence of perfect knowledge of $\bm{H}$ at the RX, one can estimate $\bm{H}$ by sending appropriate number of pilot symbols prior to the transmission of data symbols to estimate the entire or part of $\bm{H}$ as described in \cite[Sec. III-E]{Firag24}.\\
\indent To reduce the complexity, the linear model in \eqref{channel:DT:eq1} can also be used for symbol-by-symbol demodulation. In light of \eqref{channel:DT:eq1} and the knowledge of $g^{\bullet}$, the $k$th transmit symbol can be estimated using maximum likelihood (ML) detection as:
\begin{align} \label{nb:statistics:eq2}
	\hat{s}_k &= \min_{s_k \in \mathcal{Q}} \left|r(k)-\sqrt{E_s}g^{\bullet}s_k \right|,
\end{align}
for $k=0,1,\dots, L-1$, where $\mathcal{Q}$ is the constellation, which the symbols $s_k$s are drawn from.
\begin{figure}[t]
	\vspace{-5mm}
	\centerline{\includegraphics[scale=0.6]{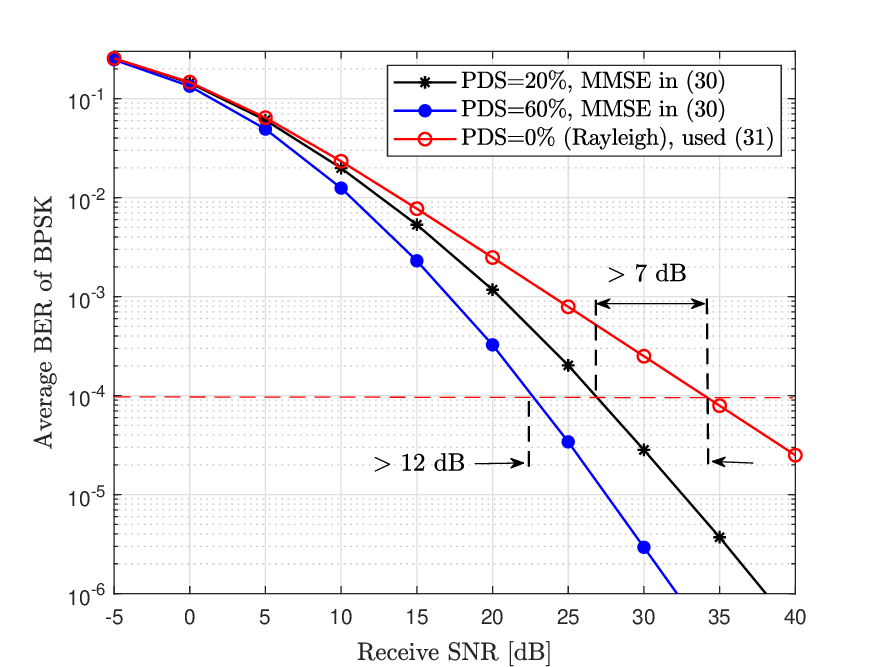}}
	\caption{\cite[Fig. 3]{Firag24} Average uncoded BER performance of BPSK modulation in a severe NLoS propagation with $\kappa=0$ for PDS$=20\%,60\%$, where $T_s=0.5\mu s$, $\beta=0.22$, and $L=40$. The span of RC/SRRC filters is set to 12 symbol periods meaning $K=6$, and the $T_m$ is chosen to satisfy PDS constraints. PDS$=0\%$ denotes to the case where all the MPCs have no relative propagation delay. The perfect knowledge of $\bm{H}$ at RX is assumed.}\label{fig:fig12}
\end{figure}
%
%\balance
%
\section{Computer Simulation}
\begin{figure}[t]
	\vspace{-5mm}
	\centerline{\includegraphics[scale=0.6]{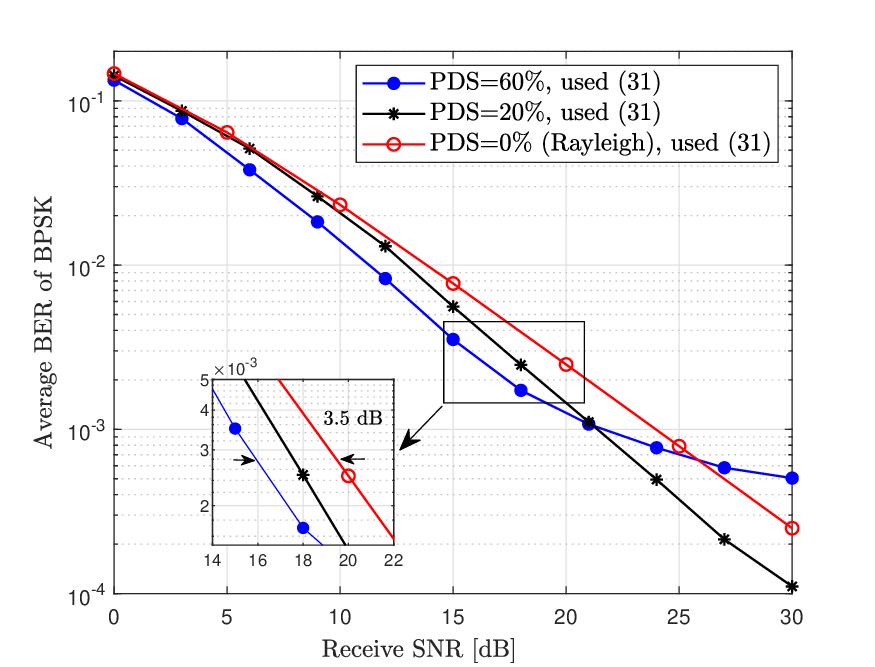}}
	\caption{Average uncoded BER performance of BPSK modulation in NLoS propagation for PDS$=20\%,60\%$, where $T_s=0.5\mu s$, $L=200$, $\kappa=0$, $K=6$. PDS$=0\%$ denotes to the case where all the MPCs have no relative propagation delay. Here the perfect knowledge of $g^{\bullet}$ at RX is assumed.}\label{fig:fig13}
\end{figure}

The starting point of our simulation is the right-hand-side (RHS) of \eqref{eq00}. Using an appropriate multipath delay profile satisfying a certain PDS, we obtain $\left\{\tau_n, \gamma_n \right\}$. After optimum symbol timing synchronization, which is equivalent to finding $\hat{\tau}_i$ and $\hat{\tau}_q$ invoking the optimizations in \eqref{eq:toi} and \eqref{eq:toq} respectively, $y(t)$ is subsequently sampled regularly at every $T_s$ interval. For this sampling, in conjunction with the definition of RC pulse in \eqref{eq2}, we use the $y(t)$ as described in \eqref{eq0}. We further assume that the constellation is BPSK, so $s_k \in \left\{-1,1\right\}$ for $\forall k$. Note that we use separate symbol timing synchronization for the I and Q branches; therefore, sampling is also performed separately for the I and Q branches, as described in \eqref{eq:samplingIQ}. In conjunction with the noise samples with appropriate variance as described in Sec. \ref{sec:SNR}, we can obtain $\bm{r}$. Subsequently exploiting a model either in \eqref{eq:bigrr} or \eqref{channel:DT:eq1}, we detect the data as outlined in Sec. \ref{sec:ber}.\\
\indent Note that the models in \eqref{eq:bigrr} and \eqref{channel:DT:eq1} should be used for demodulation only, and the foundation model in \eqref{eq00} along with optimum sampling should be used to obtain the receive samples, $\bm{r}=\left\{r(0),\dots, r(L-1)\right\}^T$. For the sake of verifiability and reproducibility, the Annexes A-D include a detailed simulation of a single-input-single-output (SISO) mediumband wireless communication system satisfying a certain PDS. The main program that estimates BER is in Annex A, and three required functions are given in Annexes B-D.\\
\indent For the sake of completeness, furthermore, Figs. \ref{fig:fig12} and \ref{fig:fig13} show the results of a representative simulation of a mediumband wireless communication system for different PDSs. The readers are advised to study these figures alongside the MATLAB codes in Annexes A-D, which are comprehensively-commented. Note that PDS$=0\%$ denotes to the case where all the MPCs have no relative propagation delays causing models in \eqref{eq:bigrr} and \eqref{channel:DT:eq1} to reduce to an one model. That happens because when PDS$=0\%$, $\mathcal{I}(k)\rightarrow 0$ $\forall k$ and $h_{\nu} \rightarrow 0$ $\forall \nu$ except $\nu=0$. When $\nu=0$, the fading coefficient: $h_{0}=h_{0}^{\bullet} \rightarrow g^{\bullet}$. Furthermore, using the programs in Annexes A-D, readers should be able to replicate and verify the results not only in Figs. \ref{fig:fig12} and \ref{fig:fig13}, but also in \cite{Bas2023,BasMag2024,Firag24}.       
%
%\balance
\section{Conclusion}
This paper briefly, but including key information, presented the necessary steps required to simulate a SISO mediumband wireless communication system with a certain degree of mediumbandness. In particular, we elaborated how optimal symbol timing synchronization and sampling should be achieved to harness the true potential of operating in the mediumband.  Furthermore, this paper includes a detailed, comprehensively-commented sample MATLAB code to simulate a mediumband wireless communication system in Annexes A-D on pages 10-12. This paper deliberately omitted a detailed description of mediumband wireless communication and its broader impact for brevity, but readers are referred to key references herein for further details in that regard.\\
\indent We hope this paper would be helpful for both experts and novices to study wireless communication systems more accurately understanding delicate effects such as the effect of deep fading avoidance, and in turn exploiting them for multiple-input-multiple-output (MIMO), cognitive radio, relaying, intelligent reflecting surfaces (IRS), space-time coding, limited feedback, non-orthogonal multiple access, integrated sensing and communication (ISAC), and countless other emerging application areas of RF wireless communication.
\appendices
\section{Proof of Matrix Model}\label{App:A}
In light of \eqref{eq0} and \eqref{eq1} and considering the fact that RC filter, $p(t)$ is limited to $2K$ symbol periods, the $l$th sample of $r(t)$ taken at $t=\hat{\tau}+lT_s$ can, in the absence of noise, be rewritten as:
\begin{align}
	\! r(l)= & \sqrt{E_s} \sum_{n=0}^{N-1} \! \gamma_n 	%    \nonumber\\
	%    &
	\left[\sum_{\nu=-K+l}^{K+l} \! \!s_{\nu} p(\hat{\tau} + lT_s\! - \!\tau_n \! - \!\nu T_s)\right],
	\label{appen:eq:1}
\end{align}
for $l=0,1,\dots, L-1$. After changing the order of the summations, \eqref{appen:eq:1} becomes:
\begin{align}
	\! r(l)= & \sqrt{E_s} \!\! \sum_{\nu =-K+l}^{K+l}  s_{\nu} \! %    \nonumber\\
	%    &
	\left[ \sum_{n=0}^{N-1}  \gamma_n  p(\hat{\tau} + lT_s \! - \!\tau_n \! - \!\nu T_s)\right].
	\label{appen:eq:2}
\end{align}
Using the fact that $s_{\nu}=0$ for $L \leq \nu < 0$, one can arrive at the matrix formation in \eqref{eq:bigrr} easily. 
%
%\balance
%
\section{Variance of Noise Samples}\label{App:B}
We are interested in the variance of the noise samples, $w(k)$ in \eqref{eq:bigrr} and \eqref{channel:DT:eq1}. Without loss of generality, we consider only the real part of $w(k)$, that is $w_i(k)$. This $w_i(k)$ in fact is the $k$th sample of the coloured noise process, $w_i(t)$ in the in-phase (I) branch, which is:
\begin{align}
	w_i(t) &= \int_{-\infty}^{\infty} \bar{w}(u) p_R(t-u) du,
\end{align}
\balance
where $\bar{w}(u)$ is the zero mean additive white Gaussian noise (AWGN) process with  power spectral density (PSD) of $N_0/2$. The autocorrelation function of $w_i(t)$ would thus be: 
\begin{align}
	\mathcal{E}\left\{w_i(t)w_i(t+\tau)\right\} &=  \nonumber \\
	& \!\!\!\!\!\!\!\!\!\!\!\!\!\!\!\!\!\!\!\!\!\! \!\!\!\!\!\!\!\!\!\!\! \!\!\!\!\!\!\!\!\!\!\! \!\!\!\! \mathcal{E}\left\{\int_{-\infty}^{\infty}\int_{-\infty}^{\infty}\bar{w}(u) \bar{w}(v)  p_R(t-u)  p_R(t+\tau-v) du dv\right\}, \nonumber \\
	&\!\!\!\!\!\!\!\!\!\!\!\!\!\!\!\!\!\!\!\!\!\! \!\!\!\!\!\!\!\!\!\!\! \!\!\!\!\!\!\!\!\!\!\! \!\!\!\! =\frac{N_0}{2}\int_{-\infty}^{\infty}\int_{-\infty}^{\infty}\delta(u-v) p_R(t-u)  p_R(t+\tau-v) du dv \nonumber \\
	&\!\!\!\!\!\!\!\!\!\!\!\!\!\!\!\!\!\!\!\!\!\! \!\!\!\!\!\!\!\!\!\!\! \!\!\!\!\!\!\!\!\!\!\!\!\!\!\! =\frac{N_0}{2}\int_{-\infty}^{\infty}p_R(t-v)  p_R(t+\tau-v)dv \label{auto_proof_eq1}
\end{align} 
The variance of $w_i(k)$, $\var\left\{w_i(k)\right\}$ is equal to $	\mathcal{E}\left\{w_i^2(t)\right\}$, which is \eqref{auto_proof_eq1} when $\tau=0$. From \eqref{auto_proof_eq1}:
\begin{align}\label{auto_proof_eq2}
	\var\left\{w_i(k)\right\} &=  \frac{N_0}{2}\int_{-\infty}^{\infty}p_R(t-v)  p_R(t-v)dv,\nonumber \\
	&= \frac{N_0}{2}\int_{-\infty}^{\infty}p_R^2(v)dv \stackrel{\Delta}{=} \frac{\sigma^2}{2}= \frac{N_0}{2T_s},
\end{align}
which confirms that $\sigma^2=N_0/T_s$. As defined in \eqref{eq201}, $p_R(v)$ and $P_R(f)$ are Fourier transform pairs, which yields:
\begin{align}
	\int_{-\infty}^{\infty} p_R^2(s) ds &= \int_{-\infty}^{\infty} P_R^2(f) df \nonumber  \\  
	&= \frac{1}{T_s} \int_{-\infty}^{\infty} P(f) df = \frac{1}{T_s},
\end{align}    
and the last integral is from \eqref{eq2n1}. One can omit the derivations here, but the result in \eqref{auto_proof_eq2} is key. Furthermore, the noise samples taken $T_s$ apart are uncorrelated.
%
%\balance
%

%
\renewcommand{\appendixname}{Annex}
\appendices 
\begin{figure*}[b]
\section{Main Mediumband BER Program}\label{App:P1}
\begin{lstlisting}[style=Matlab-Pyglike]
Ts=5e-07; %This symbol Period corresponds to a 2MHz data signal.
beta=0.22; %Roll-off factor for RC and SRRC filters. Any value in [0,1].
kappa=0; %This is to shape the power-delay profile. kappa=0 means all MPCs 
%are on average equally strong.
K=6; %This limits the RC/SRRC filter span to 2K number of symbol periods.
PDS=60; %PDS as a percentage as defined in (14)
Tm=(PDS/100)*Ts; %Setting the delay spread. If one wants to generate ISI free 
%narrowband channels, simply make PDS or Tm equal to approximately zero.
Es=1;
%=========================================
%Setting SNR for BER evaluation. See Sec. III-C
SNRdB=[0:3:21]; %This ensures program repeats 8 times at SNRdB=0,3,6,...,21.
SNR=10.^(SNRdB./10);
sigma2=1./SNR;
%=========================================
L=250; %Length of PHY frame in terms of symbol periods
PL=50;	%Length of pilots in terms of symbol periods
N=10; %Setting the number of multipath components (MPC).
nsim=5000; %The number of iteration that the program runs per SNR value.
BER=zeros(1,length(SNR)); %To hold BER values.
for c=1:length(SNR)
   error_count=zeros(1,nsim);
   for i=1:nsim
      %=========================================
      %Generating BPSK samples, which means equally likely 1s and -1s,
      sp=sign(rand(1,PL)-0.5);  %for pilot symbols, and
      sd=sign(rand(1,L-PL)-0.5); %for data symbols.
      s=sqrt(Es)*[sp sd]; %Here Es is unity.
      %=========================================
      %NLOS Model
      tau0=Tm*rand(1,N); %Drawn from a uniform distribution: U(0,Tm).
      tau0=tau0-min(tau0);
      tau0=(Tm/max(tau0))*tau0;
      tau=sort(tau0); %By making all path delays equal to some arbitrary value
      %would create an ideal ISI-free narrowband channels, that is PDS=0%.
      alpha=exp(-kappa.*(1:N)); %This is just to emulate exponential power delay
      %profile of MPCs. kappa=0 means all MPCs are on average equally strong.
      alpha=(1/sqrt(sum(alpha.^2)))*alpha; %This is to ensure that the sum
      %of the average powers of MPCs is equal to unity, that is Gamma=1.  
      gamma=sqrt((alpha.^2)/2).*(randn(1,N)+j*randn(1,N)); %This is to generate
      %random-strength MPCs with random phase, but on average equal to alpha.
      %Note that one may use any application specific model to generate these gammas
      %and taus, if needed.
      %==========================================
      %This block is to find timing offsets: tauIhat and tauAhat for I and Q 
      %branches respectively. See ANNEX C for getOffset(.) function.
      [tauIhat,~]=getOffset(real(gamma),tau,beta,Ts);  %Do the optimization in (17)
      [tauQhat,~]=getOffset(imag(gamma),tau,beta,Ts);  %Do the optimization in (18)
      TI=tauIhat:Ts:(L-1)*Ts+tauIhat;
      TQ=tauQhat:Ts:(L-1)*Ts+tauQhat;
      yI=zeros(N,length(TI));
      yQ=zeros(N,length(TQ));
      y=zeros(1,length(yI));
      %Go to next page
\end{lstlisting}
\end{figure*}
\begin{figure*}[t]
	\vspace{-10mm}
	Cont.
\begin{lstlisting}[style=Matlab-Pyglike]
      %From previous page
      %==========================================
      %Evaluate g. See ANNEX D for getAutoCorr(.) function.
      gI=sum(real(gamma).*getAutoCorr(tau-tauIhat,beta,Ts)); 
      gQ=sum(imag(gamma).*getAutoCorr(tau-tauQhat,beta,Ts)); 
      g=(1/(1-0.25*beta)).*(gI+j.*gQ);   %From (26) and (27).
      %==========================================
      %This block is to sample the receive signal at every Ts starting from
      %t=tauIhat and t=tauQhat for I and Q branches respectively.
      for k=1:N
         for i=1:L
            yI(k,:)=yI(k,:)+s(i)*getRCFilter(TI-(i-1)*Ts-tau(k),beta,Ts);
            yQ(k,:)=yQ(k,:)+s(i)*getRCFilter(TQ-(i-1)*Ts-tau(k),beta,Ts);
         end
         yI(k,:)=real(gamma(k))*yI(k,:);
         yQ(k,:)=imag(gamma(k))*yQ(k,:);
      end
      y=sum(yI)+j*sum(yQ);
      %=========================================
      %This is for symbol-by-symbol detection for BPSK using model in (24)
      noise=sqrt(0.5*sigma2(c))*(randn(1,L)+j.*randn(1,L));
      r=y+noise;
      re=r(1:PL);
      ghat=(1/PL)*(re*sp'); %Simple LS estimation.
      w=g; %If one wishes to use estimated CSI, they may use ghat instead.
      rd=r(PL+1:1:end);
      rdhat=sign(real((1/(w'*w))*w'*rd));
      err_count(i)=sum(sign(abs(rdhat-sd))); %Invoking (32)
   end
%Calculate the average BER of BPSK modulation.
BER(c)=sum(err_count)./(nsim*(L-PL));
end
%Plotting
figure(1)
semilogy(SNRdB,BER,'r-*')
\end{lstlisting}
\hrule
\end{figure*}
\begin{figure*}[t]
\vspace{-0mm}
\section{$\text{getRCFilter()}$ Function}\label{App:P3}
\begin{lstlisting}[style=Matlab-Pyglike]
function [out] = getRCFilter(t,beta,Ts) %As defined in (5)
   out=zeros(1,length(t));
   out=sinc(t./Ts).*(cos(pi*t*beta/Ts)./(1-((2*beta*t)/Ts).^2));
   %If one wants to limit the span of RC filter, the following line limits the RC
   %filter span to 2K number of symbol periods.
   out(abs(t)>K*Ts)=0;
end
\end{lstlisting}
\end{figure*}
\begin{figure*}[t]
\vspace{-55mm}
\section{$\text{getOffset()}$ Function}\label{App:P2}
	\begin{lstlisting}[style=Matlab-Pyglike]
function [opt_delay,opt_amp] = getOffset(gamma,tau,beta,Ts)
   %This function performs the optimization in (17)/(18)
   %It outputs optimum time offsets (tauIhat or tauQhat) at RX.
   %It does nothing but exhaustively searches the excess delay axis
   %to find the optimum time instance that gives the strongest g. 
   N=length(tau);
   upsample_val=1207; %If the upsample_val is increased, the
   %accuracy of tauIhat and tauQhat estimation would increase.
   Tc=Ts/upsample_val;
   tp=-2*Ts:Tc:3*Ts; %We search the excess delay axis in [-2Ts, 3Ts], which is the 
   %most likely region for the maximum to exist.
   %=================================================
   xp=zeros(1,length(tp));
   for k=1:N   %This evaluates the objective function in (17)/(18)
      xp=xp+real(gamma(k))*getAutoCorr(tau(k)-tp,beta,Ts);
   end
   xxp=(1/(1-0.25*beta)).*(xp);
   xxpN=abs(xxp).^2;
   %=================================================
   [pks,locs]=findpeaks(xxpN);
   [~,Ipks]=max(pks);
   opt_delay=tp(locs(Ipks));
   opt_amp=xp(locs(Ipks));
end
	\end{lstlisting}
\hrule
\end{figure*}
\begin{figure*}[t]
\vspace{-120mm}
\section{$\text{getAutoCorr()}$ Function}\label{App:P4}
	\begin{lstlisting}[style=Matlab-Pyglike]
function [out] = getAutoCorr(tau,beta,Ts)
   %As defined in (16), this function calculates Rss(tau). 
   out=(sinc(tau./Ts).*(cos(tau*beta*pi/Ts)./(1-((2*beta*tau)/Ts).^2)) - 
       (beta/4)*(sinc(beta*tau./Ts).*(cos(tau*pi/Ts)./(1-((beta*tau)/Ts).^2))));
end		
	\end{lstlisting}
\hrule
\end{figure*}

\begin{thebibliography}{00}
	%
	\bibitem{Bas2023}
	D. A. Basnayaka, ``Introduction to mediumband wireless communication,'' \textit{IEEE Open Journal of the Commun. Society}, vol. 4, May. 2023.
	%
	\bibitem{BasMag2024}
	D. A. Basnayaka, ``Communicating in the mediumband: What it is and why it matters,'' \textit{in IEEE Communication Magazine}, Nov. 2024.
	%
	\bibitem{Ungerboeck82}
	G. Ungerboeck, ``Channel coding with multilevel/phase signals,'' IEEE Transactions on Information Theory, vol. 28, no. 1, pp. 55-67, Jan. 1982.
	%
	\bibitem{Robert19}
	R. A. Heath Jr. and A. Lozano, ``Foundations of MIMO Communication,'' Cambridge University Press 2019.
	%
	\bibitem{BasFirag25}
	D. A. Basnayaka, and A. Firag, ``Integrating mediumband with emerging technologies: Unified vision for 6G and beyond physical layer,'' Online: ArXiv:https://arxiv.org/abs/2501.10122.
	%
	\bibitem{Gold05}
	A. Goldsmith, Wireless Communications, Cambridge Uni. Press, 2005.
	%
	\bibitem{Ezio07}
	E. Biglieri and et. al., ``MIMO Wireless
	Communications,'' Cambridge University Press 2007.
	%
	\bibitem{Firag24}
	A. Firag, J. Jia and D. A. Basnayaka, ``A link-level performance analysis for mediumband wireless communication,'' \textit{IEEE Virtual conference on Communications}, Dec. 2024.
	%
\end{thebibliography}
\end{document}